\documentclass[runningheads]{llncs}
\usepackage{graphicx}
\usepackage{array}
\usepackage{url}
\usepackage{tabularx} 
\usepackage{multirow}
\usepackage{color}
\usepackage{numprint} 
\npthousandsep{,}
\usepackage{arydshln} 
\usepackage{balance} 
\usepackage{todonotes}

\usepackage{framed}

\usepackage{numprint}
\npthousandsep{,}

\begin{document}
\title{Evaluating Maintainability Prejudices\\ with a Large-Scale Study of Open-Source Projects}
\titlerunning{Evaluating Maintainability Prejudices}
%
\author{Tobias Roehm\inst{1} \and
Daniel Veihelmann\inst{1} \and
Stefan Wagner\inst{2}\orcidID{0000-0002-5256-842} \and
Elmar Juergens\inst{1}}

\authorrunning{T. Roehm et al.}
%
\institute{CQSE GmbH, Munich, Germany
\email{roehm,veihelmann,juergens@cqse.deu} \and
University of Stuttgart, Germany
\email{stefan.wagner@iste.uni-stuttgart.de}}
\maketitle              
\begin{abstract}


Exaggeration or context changes can render maintainability experience into prejudice. For example, JavaScript is often seen as least elegant language and hence of lowest maintainability. Such prejudice should not guide decisions without prior empirical validation.

We formulated 10 hypotheses about maintainability based on prejudices and test them in a large set of open-source projects (6,897 GitHub repositories, 402 million lines, 5 programming languages). We operationalize maintainability with five static analysis metrics.

We found that JavaScript code is not worse than other code, Java code shows higher maintainability than C\# code and C code has longer methods than other code. The quality of interface documentation is better in Java code than in other code. Code developed by teams is not of higher and large code bases not of lower maintainability. Projects with high maintainability are not more popular or more often forked. Overall, most hypotheses are not supported by open-source data.

\end{abstract}
\section{Introduction}
\label{sec:introduction}

To implement a feature or fix a bug, software developers have to identify relevant code regions and comprehend them. As program comprehension is part of many maintenance tasks, developers spent 50--70\% of their time on it~\cite{fjeldstad1983application,minelli2015know}. Source code with high quality makes it easier for developers to comprehend it~\cite{martin2009clean}. Hence, developers working with highly maintainable code can fix bugs faster, implement features more rapidly, and spend less effort compared to developers working with low-quality code.

Due to this importance of maintainability, developers form their own ``theories'' of the
interrelations of programming languages, team characteristics or code size with maintainability based on their daily experiences. By cumulating and discussing such experiences with others, they can become ``prejudices''. Such prejudices will then be the basis for decisions in projects.
Therefore, empirical research should evaluate whether such prejudices are really supported by facts and data. Yet, there are few empirical studies 
on these subjects. 

To help close this gap, we contribute a large-scale study about ten hypotheses derived from our practical experience with prejudices related to maintainability in
open-source code. We investigate maintainability in relation to programming languages, project size and project popularity. 
We operationalize the abstract concept of maintainability using five static analysis metrics: clone coverage, too long files, too long methods, nesting 
depth and comment incompleteness. All these metrics identify code smells which increase the effort for developers during program comprehension.
While they cannot cover maintainability completely, based on existing studies and our consulting experience, they are good indicators for maintainability.

We chose GitHub as a source of open-source code, randomly selected repositories, downloaded their code, excluded irrelevant code automatically and performed static analysis. We considered GitHub repositories with code written in C, C\#, C++, Java, or JavaScript. We chose these languages because (1) they are all in the C-family of languages, which makes
our used metrics better comparable, and (2) our used analysis tool covers them well. Overall, our data set contains \numprint{6,897} repositories containing 402 million source lines of code.\footnote{``Lines of code" denotes all lines in a file or method, ``source lines of code" all lines while ignoring empty lines and comments} 

This paper complements other empirical studies on quality, especially the studies by Ray et al.~\cite{Ray2014large,Ray2017} and Bissyand\'e et al.~\cite{bissyande2013popularity}. 
They operationalize quality focusing on the number of bug fixing commits or the number of issues. While faults are an interesting aspect of quality, we found that our set of static code metrics represent the maintainability side.

The contribution of this paper is a large-scale study of maintainability in open-source projects. The study investigates 10 prejudices
formalized as statistical hypotheses. It covers the comparison of the maintainability of code in different languages overall as well as specific aspects
of maintainability. Furthermore, we investigate the relationships of maintainability and the team and code size as well as the forks and popularity of open-source
projects.
To the best of our knowledge, this is the largest study of maintainability and the first large-scale study of maintainability in open-source projects operationalizing maintainability with a set of static analysis metrics. We provide a 
replication package.\footnote{see https://github.com/Dan1ve/MSR17CodeQualityOnGitHub} 


\section{Related Work}
\label{sec:related-work}



Ray et al.~\cite{Ray2014large,Ray2017} present a similar study on code quality of GitHub code. They investigate the effect of programming languages and their properties as well as the impact of application domains on code quality. In contrast to this study, they operationalize code quality by the number of bug fix commits. Bissyand\'e et al.~\cite{bissyande2013popularity} present another similar study on software quality of GitHub repositories. They investigate the impact of programming languages on project success, code quality, and team size. In contrast to this study, they measure code quality by the number of issue reports. We offer the complementary view of static analysis.

Static analysis has been used by other researchers to investigate the quality of open-source code. But most of that research considers few software applications or few programming languages. Samoladas et al.~\cite{samoladas2004open} and Stamelos et al.~\cite{stamelos2002code} present studies on code quality of open-source code. Norick et al.~\cite{norick2010effects} present a study on open-source code, investigating the impact of team size on code quality. All three papers also use static analysis to measure code quality, they use different metrics, which we consider less suitable for maintainability. Furthermore, they do not investigate the impact of programming language or other factors on quality.  Ahmed, Ghorashi and Jensen~\cite{ahmed2014exploration} investigate the code quality of open-source code by examining the correlation between code smells and project characteristics. They do not investigate the impact of programming language and use different static analysis metrics. 

All these studies either used outdated and strongly criticised metrics such as the Halstead metrics, the Maintainability Index or more detailed bug pattern analysis. We
concentrated on metrics that are automatically collectable, have been proven in practice to be good indicators for quality and are measurable across languages
\cite{ostberg2014automatically,steidl2014continuous,heinemann2014teamscale}.
Only Koschke and Bazrafshan~\cite{koschke2016software} also investigate cloning in their study on clone rates in code written in C and C++. Yet, they focus only on programming 
languages and cloning.

Several studies investigated aspects of software quality. Kochhar, Wijedasa and Lo~\cite{Kochhar2016large} present a study on code quality of GitHub code which examines the impact of language co-use on code quality. Bird et al.~\cite{bird2009does} did a study on software quality of industry code which investigates the influence of distributed development on software quality. Bird et al.~\cite{bird2011dont} also present a study on software quality of industry code which studies the impact of code ownership on software quality. Nagappan, Murphy and Basili~\cite{nagappan2008influence} examine software quality of industry software by investigating the effect of organizational structure on software quality. All these studies consider different aspects of software quality than this study and operationalize quality differently.

For a more detailed overview of code quality measurement of open-source code we refer to Ruiz and Robinson~\cite{ruiz2013measuring} and Spinellis et al.~\cite{spinellis2009evaluating}. Beller et al.~\cite{Beller2016analyzing} present a study on static analysis tools on GitHub. While they focus on the usage of static analysis, we take advantage of static analysis results as an operationalization of code quality.

\section{Study Design}
\label{sec:study-design}


\subsection{Research Questions}

The focus of this study is prejudices on maintainability. We investigate the following two research questions:


\noindent\textbf{RQ1: How does the programming language affect maintainability?}
One major area of prejudices is about the impact of the programming language on maintainability. We examine whether code written in one programming language differs in quality from code written in another programming language. We formulate six hypotheses about potential relationships between programming language and maintainability.

\noindent\textbf{RQ2: How do non-language aspects influence or are influenced by maintainability?}
In addition to the programming language, many other aspects are often seen as factors that might influence maintainability. This study investigates further factors by using the meta data of GitHub repositories and relating them to maintainability metrics. The following aspects form four more hypotheses to be tested: 
code base size, team size, individual vs. team code, development activity and repository popularity.
%


\subsection{Hypotheses Formalizing Prejudices}
\label{sec:hypotheses}

To make prejudices analyzable in an empirical study, we formulated ten hypotheses about maintainability before performing data collection and analysis (cf.~Table~\ref{tab:hypotheses-overview}). This avoids cherry-picking obvious results from the dataset and thereby ``overfitting" to the studied dataset~\cite{eisenhardt1989building}. Our hypotheses can be divided in two categories: assumptions about the impact of the programming language on maintainability\footnote{Apart from personal experience and discussions based on blog posts such as \url{http://live.julik.nl/2013/05/javascript-is-shit}} (cf.~RQ1) and hypotheses about the influence of other aspects on maintainability (cf.~RQ2). The hypotheses consider the languages C, C++, C\#, Java, and JavaScript.

The motivation behind H1 and H4 is the fact that C is not an object-oriented language. Hence, in situations where developers of object-oriented languages can use inheritance, developers of C code probably have to duplicate the code. H2 was derived from the fact that Java and C\# are very similar languages. The idea behind H3 is that JavaScript code might contain anonymous callback functions more frequently, which might lead to deeper nested code. Since documentation frameworks like JavaDoc are available for all of the studied languages, we assume that there is no difference in this regard (H5). The motivation behind H6 is the assumption that JavaScript development often has a rapid development pace, making it challenging to produce high-quality code. H7 is based on the prejudice that team members push each other to develop high-quality code and high-quality code is necessary for collective code ownership. The motivation behind H8 is the assumption that small code bases can keep higher maintainability standards more easily than large code bases. The idea behind H9 is the assumption that developers like to work with high-quality code bases and, hence, such repositories have more forks.

\begin{table}
\caption[Overview of Hypotheses]{Overview of Hypotheses\\``Other languages" and ``All languages" refers to the set (C, C++, C\#, Java, JavaScript).}
\label{tab:hypotheses-overview}
\begin{tabularx}{\columnwidth}{ l X }
\hline
\multicolumn{2}{l}{Hypotheses on Languages} \\
\hline
  H1 & C code has more code duplication than code written in other languages.\\
  H2 & Code written in Java and C\# has no differences regarding maintainability.\\
  H3 & JavaScript code is more deeply nested than code written in other languages.\\
  H4 & C code has more very long methods than code written in other languages.\\
  H5 & The quality of interface documentation is similar for all languages.\\
  H6 & JavaScript code has the lowest maintainability among all languages.\\
\hline
\multicolumn{2}{l}{Hypotheses on Project Size} \\
\hline 
  H7 & Code developed by a team has a better quality than code developed by an individual.\\
  H8 & Large code bases have a lower quality than small code bases.\\
\hline
\multicolumn{2}{l}{Hypotheses on Project Popularity} \\
\hline 
  H9 & Repositories with high maintainability have more forks than repositories with low maintainability.\\
  H10 & Repositories with high maintainability are more popular than repositories with low maintainability.\\
\hline
\end{tabularx}
\end{table}


\subsection{Object Selection}
\label{sec:case-and-subjects-selection}

This study uses GitHub~\cite{github-homepage} as source of open-source code because the platform hosts a huge number of repositories and it provides infrastructure for selecting and downloading code (see Gousios~\cite{Gousios2013ghtorrent} and the GitHub API~\cite{github_api-homepage}). We used the GHTorrent data set~\cite{Gousios2013ghtorrent} (Version 2016-03-16) to pre-select relevant GitHub repositories. GHTorrent is a repository of GitHub meta data.
\vspace{-1em}
\subsubsection*{Inclusion \& Exclusion Criteria}
We defined inclusion and exclusion criteria that a GitHub repository has to fulfill to be relevant for this study. The goal of these criteria is to maximize the validity of the results. Some criteria are based on advice by Kalliamvakou et al.~\cite{Kalliamvakou2014promises}. We applied the following criteria:

\emph{Programming Language}: One of the repository's languages has to be C, C++, C\#, Java, or JavaScript. We focus on these languages because they are all in the C-family and, hence,
the metrics are more likely to be comparable across languages, the used code analysers support these languages, and we reuse preliminary work when excluding irrelevant code (cf. Section~\ref{sec:data-collection-procedures}).

\emph{Minimum Size of Code Base}: The code base must have at least \numprint{10000} lines of code. This criterion is used to exclude small repositories which might bias the results and are likely to be dummy repositories.

\emph{No Fork or Mirror}: A repository must not be a fork or mirror of another repository. This criterion is used to concentrate on the main repositories and avoid analyzing the same code base multiple times. 

\emph{Availability of Description}: The repository must have a readme file and/or a GitHub description. This criterion is used to filter out dummy repositories.

\emph{Maximum Size of Code Base}: The size of the code base per language must not exceed 215 MB. This criterion is used to exclude large repositories which contain the same code multiple times, e.g. in different versions. 99\% of repositories in the GHTorrent dataset are below this threshold and an empirical investigation (conducted by two authors as part of this study) showed that 75\% of repositories exceeding this threshold contain the same code multiple times. 

\emph{Maximum Clone Coverage}: The clone coverage of a repository must be smaller than 75\%. When manually analyzing code bases with high clone coverage, we found that the majority of them contained the same application multiple times. Empirical evaluation of different threshold values revealed that a threshold of 75\% is a good compromise between considering as many repositories as possible and excluding irrelevant repositories.   

When applicable, these criteria were applied to the GHTorrent data set. The criteria ``Minimum Size of Code Base" and ``Maximum Clone Coverage" were applied after static analysis because they require the corresponding results. Only repositories that fulfill all criteria were added to the final data set and used in further analysis. The final data set consists of \numprint{6897} repositories containing 401 million source lines of code.
\vspace{-1.2em}
\subsubsection*{Random Sampling}
The GitHub repositories considered in the study were selected randomly. To perform this step, the list of GitHub repositories matching the criteria was ordered randomly and repositories were downloaded starting at the top of the list. 

\subsection{Data Collection Procedures}
\label{sec:data-collection-procedures}

After selecting the GitHub repositories, we downloaded their code and removed generated code, test code and library code to focus the analysis on production code and improve comparability of results.
\vspace{-1.4em}
\subsubsection*{Download via GitHub API}
The code of selected GitHub repositories was downloaded using the GitHub API~\cite{github_api-homepage}. In addition to code, we retrieved meta data of repositories like the number of committers or the number of stars. This meta data is used to investigate the impact of non-language aspects on maintainability (cf. RQ2).
\vspace{-1.4em}
\subsubsection*{Exclusion of Irrelevant Code}

Code generated by tools distorts the results of static analysis. This is especially true for clone detection as generated code often follows a pattern which is likely to be classified as a clone. Hence, we exclude generated code. Test code is code executed to verify the correctness of an application~\cite{beck2003test}. Despite its importance, we exclude test code to achieve better comparability between repositories and because developers might not apply high quality standards to it (c.f.~Steidl and Deissenboeck~\cite{steidl2015java}). When manually analyzing a sample of GitHub repositories, we found that repositories often contain library code. Library code denotes code that is developed by a third party and was copied into a repository, probably for reuse purposes. This is  common for JavaScript repositories where library code is often included as source files and not as binaries. We exclude library code since we solely focus on the primary repository code.

We used four mechanisms to exclude irrelevant code: 

\textbf{Comment exclusion} uses code comments that indicate that a file is generated or part of a library, e.g.\  ``@license AngularJS v1.0.7 (c)". We use a list of \numprint{2247} exclusion comments assembled by Hoenick~\cite{Hoenick2015does}.  According to Hoenick~\cite{Hoenick2015does}, this approach excludes 97\% of generated code. 

\textbf{Path exclusion} is based on file system paths which indicate that code files in a directory are generated, test code or library code, e.g.\ ``**/generated/**". We use a list of \numprint{58} exclusion paths assembled by Hoenick~\cite{Hoenick2015does}. 

\textbf{Import exclusion} exploits import statements of popular test frameworks such as JUnit. We use a list of 10 test framework imports assembled by Hoenick. According to him, this approach excludes 90\% of test code. 

\textbf{File name exclusion} is based on the file name frequency of popular library files. We counted the frequency of file names in all downloaded repositories, manually reviewed the 200 most frequent file names, extracted a list of 60 library files, and excluded the corresponding files. Furthermore, we ignore minified JavaScript -- code which was automatically shrunk to reduce file size and thus looks significantly different than the original code -- by its file suffixes, e.g.\ ``min.js". These efforts aim at excluding library code as precisely as possible. A manual evaluation indicated that we got rid of most library code. 



\subsection{Operationalization of Maintainability}
\label{sec:operationalization-of-code-quality}

The operationalization of software quality in general and also maintainability in particular has not been solved satisfyingly in general. Several
quality model approaches have aimed at systematically deriving good indicators\cite{wagner2015operationalised,samoladas2008sqo,heitlager2007practical}. Yet, it is difficult to cover all aspects of maintainability.
We follow here our proposal of an activity-based maintainability model \cite{deissenboeck2007activity} and focus on statically measurable indicators to be feasible for our large-scale
study. We select static metrics which we expect to have an impact on the main maintenance-related activity \emph{code comprehension} 
(cf.\ Table~\ref{tab:metrics-overview}): clone coverage, too long files, too long methods, nesting depth and comment incompleteness. Furthermore,
we chose these metrics because they are easy to understand and improve, they are language-independent, they have been found to be suitable 
for making solid statements about software maintainability~\cite{ostberg2014automatically} and they are used in practice~\cite{steidl2014continuous,heinemann2014teamscale}.

\begin{table}
\caption[Overview of Static Analysis Metrics]{Overview of Static Analysis Metrics\\ Unit of all metrics is percentage where higher values indicate lower quality.}
\label{tab:metrics-overview}
\begin{tabularx}{\columnwidth}{ p{2.5cm} X }
\hline
  Metric & Definition\\
\hline
  Clone coverage & The fraction of source lines in the code base which are part of at least one (type 2) clone.\\
	Comment incompleteness & The fraction of public classes, types, methods, procedures, and attributes which are not documented by a comment.\\
	Too long files & The fraction of source lines in the code base which are located in files exceeding 750 source lines.\\
	Too long methods & The fraction of source lines in the code base which are located in methods exceeding 75 source lines.\\
	Nesting depth & The fraction of source lines in the code base which are located in methods with at least one line exceeding nesting depth 5.\\
\hline
\end{tabularx}
\end{table}

The metric ``clone coverage"~\cite{steidl2014continuous} indicates the fraction of the code base which is part of at least one clone. If the value is high, this means that developers frequently encounter duplicated code. Code clones unnecessarily increase a code base. Furthermore, faults fixed in one clone instance can remain present in other clone instances and inconsistent clones likely introduce bugs~\cite{juergens2011why}. The metric ``too long files"~\cite{steidl2014continuous} identifies the fraction of the code base which is located in long files. Long files are often difficult to comprehend as one has to consider a large fragment of code. In addition, long files might be a hint for bad modularization. 

The metric ``too long methods"~\cite{steidl2014continuous} identifies the fraction of the code base which is located in long methods. Long methods are difficult for developers to comprehend because they have to consider much code. Furthermore, long methods might be an indicator for bad modularization. The metric ``nesting depth"~\cite{steidl2014continuous} identifies the fraction of the code base which is located in methods which are deeply nested. These are difficult to comprehend because each condition ``controlling" a nested statement has to be taken into consideration to tell when the statement is executed. 

Finally, metric ``comment incompleteness"~\cite{heinemann2014teamscale} identifies code entities like methods, classes, or attributes, which lack any kind of explanatory comment. Missing documentation makes it expensive for developers to comprehend what a code entity does and how it can be reused. We make two restrictions here: we consider only public code entities and we ignore trivial getter/setter methods as well as overriding methods. We only regard public code entities because they are open for reuse and hence we expect them to be documented. Ignoring getter and setter methods accommodates the fact that these methods are often too trivial to document. We ignore overriding methods because usually the documentation of the overridden method is sufficient.

All of these metrics use percentage values as unit of measurement where high percentage values indicate low maintainability. This makes it easier to interpret and compare metric values.

\subsection{Analysis Procedures}
We performed three types of analysis: static analysis to calculate metric values, descriptive statistics for an overview of the data and inferential statistics to test hypotheses. The 
details of these analyses are described in the following paragraphs. All statistical analyses were performed with R.
\vspace{-1.4em}
\subsubsection{Static Analysis}
\label{sec:static-code-analysis}

We use the open-source tool ConQAT~\cite{deissenboeck2008tool,conqat-homepage} to calculate static analysis metrics for each GitHub repository after download and exclusion of irrelevant code. 
We used the following static analysis parameters: For the metric ``clone coverage", we consider clones that consist of ten or more consecutive statements. In addition to identical fragments of source code, we consider clones that contain simple modifications such as variable renamings (i.e.\ type 2 clones). For the metric ``too long files", we consider source lines in files with more than 750 source lines of code in the percentage value, which is less strict than e.g. Martin~\cite{martin2009clean} who advocates file size of less than 500 lines. 
For the metric ``too long methods", we consider source lines in methods with more than 75 source lines of code in the percentage value, which is less strict than e.g.\ Martin~\cite{martin2009clean} who advocates method with less than 20 lines. While it might look inappropriate to use the same threshold values for different languages on a first glance, we argue that -- given the rather lenient threshold values where a method already spans several screens -- the constraints of developers' working memory are the dominating factor~\cite{miller1956magical}, independent of the language. 

For the metric ``nesting depth", we consider code lines in methods which have at least one statement with nesting depth 5 or deeper in the percentage value. Again, this is less strict than Martin \cite{martin2009clean} who advocates nesting depths below 2. And finally for the metric ``comment incompleteness", we consider all public types, methods, functions, procedures, properties, attributes and declarations, but we exclude simple getter methods, setter methods and override methods in the percentage value. This is in accordance with respective guidelines \cite{google-guidelines,heinemann2014teamscale}. We did not evaluate comment incompleteness for JavaScript code because this analysis is not supported by ConQAT. All configuration details can also be found in a configuration file in the replication package.
\vspace{-1.4em}
\subsubsection{Statistical Analysis}
To aggregate results from individual code bases to groups of code bases and compare results of different groups, we used basic statistics like minimum, maximum, or median. Because no data attribute exhibits a normal distribution and variances differ, we report median instead of mean 
and use corresponding statistical procedures. Nevertheless, we use MANOVA analyses for first tests if the null hypotheses could be rejected at all in cases where quality overall, and
hence all quality metrics, is involved. We believe this is a valid approach, because for more detailed comparisons, we then choose more robust non-parametric methods. Yet, if a parametric
approach finds no significant difference, it is not necessary to investigate further.

We studied ten hypotheses about maintainability (cf.\ Table~\ref{tab:hypotheses-overview}). Because they either formulate a hypothesis about a single quality metric or quality overall.
For single quality metrics, we use the Kruskal test first to see if there are any differences at all. To investigate which group is significantly different from another, we applied a (pairwise) Mann–-Whitney U test (also known as Wilcoxon rank sum test, see Kabacoff~\cite{kabacoff2015r}). In the case of pairwise tests, probability adjustment according to Holm is used. This non-parametric test can be used for two independent groups that are not normally distributed and with different group sizes. In cases where quality overall is part of a hypothesis, we use a MANOVA
first to test whether there is a difference for any quality metric. In case there is a difference, we use single ANOVA tests to see which quality metrics are significantly different. For those,
we then use the non-parametric Mann-Whitney U test for pairwise comparisons. For H8, we employ Pearson's coefficient to quantify the relationship between the size of the
code bases with the quality metrics. We use a significance level of p $<$ 0.01. We use Cohen's d for effect sizes and Cliff's delta or Pearson's
correlation coefficient as alternative where necessary.

%

\subsection{Validity Procedures}

We employed the following procedures to maximize the validity of the results. First, we refined the inclusion and exclusion criteria several times to filter out GitHub repositories which might distort the results (cf. Section~\ref{sec:case-and-subjects-selection}). Second, we used only data for which we are confident of their quality and validity. For instance, we refrained from analysing maintainability with the number of open issues, number of pull request or release count. While these would be interesting to analyze, only a fraction of GitHub repositories uses these GitHub features~\cite{Kalliamvakou2014promises}. Hence, these data attributes are probably not valid and would lead to meaningless results. Third, we manually browsed through a sample of about 50 repositories to see what kinds of artefacts are present. This analysis led to the detection and exclusion of library code and minified JavaScript code. Fourth, we manually verified the static analysis results of a sample of roughly 30 repositories. Fifth, we manually analyzed a sample of approx.\ 90 repositories with exceptionally high or low metric values. For repositories with very high clone coverage, we found that this is often caused by multiple project copies within one repository. Thus, we established a maximum value of 75\% for clone coverage. Sixth, the static analysis was mainly performed by one author and the results were reviewed by another author for validity and coherence. Seventh, we considered a large sample of repositories --- more than 600 per programming language --- in the analysis. Finally, we use a conservative significance level for the statistical test ($\alpha= 0.01$).

\subsection{Study Objects}
\label{sec:study-subjects}

Table~\ref{tab:study-subjects} provides an overview of studied GitHub repositories. Overall, about 1 million repositories in the GHTorrent data set fulfilled the selection criteria. After download and code exclusions, our data set consisted of \numprint{6897} repositories containing 402 million source lines of code overall. Only repositories that fulfill all criteria (cf. Section~\ref{sec:case-and-subjects-selection}) were added to this data set. The size of the repositories varies from 991 SLoC to 1.3m SLoC with a median size of 20k SLoC (please note that we put the minimum size constraint on LoC and not SLoC). The number of repositories per language varies because we downloaded repositories in a round robin fashion but excluded already downloaded repositories when their size fell below 10 kLoC after code exclusions. About half of the repositories (\numprint{3434}) were individual repositories, i.e.\ repositories with just one committer, while the other half (\numprint{3463}) were team repositories. The number of commits varied from \numprint{1} to \numprint{507000} with a median of \numprint{40} commits. The most forked repository had \numprint{8790} forks while most repositories were not forked at all. Similarly, the most popular repository had \numprint{19200} stars while most repositories had no stars. 

\begin{table*}
  \centering
\caption[Overview of Study Subjects/ GitHub Repositories]{Overview of Study Subjects/ GitHub Repositories\\ k = \numprint{1000}, m = \numprint{1000000}, Format of multi-value cells: Median (Min-Max)}
\label{tab:study-subjects}
\begin{tabular}{l | r r r | r r r r r c c c c c}
  Lang. & Relev. & Used  & Used & Size (SLoC) & \#com- & \#commits & \#forks & \#stars\\
        & repos  & repos & SLoC &             & mitter &           &         &         \\
  \hline
	C & 139k & \numprint{2072} & 138m & 25k (2k-939k) & 2 (1-835) & 37 (1-114k) & \numprint{0} (0-6.7k) & \numprint{0} (0-19.2k) \\
		
	C++ & 141k & \numprint{2035} & 173m & 25k (2k-1.2m) & 2 (1-15k) & 46 (1-507k) & \numprint{0} (0-8.7k) & \numprint{0} (0-15.8k) \\
	
	C\# & 70k & \numprint{716} & 22m & 15k (5k-904k) & 2 (1-110) & 55 (1-27k) & \numprint{1} (0-1.6k) & \numprint{0} (0-8.2k) \\
	
	Java & 203k & \numprint{978} & 31m & 14k (1k-710k) & 2 (1-132) & 53 (1-22k) & \numprint{0} (0-2.1k) & \numprint{0} (0-3.5k) \\

	JS & 496k & \numprint{1096} & 37m & 22k (4k-628k) & 1 (1-325) & 25 (1-119k) & \numprint{0} (0-1.9k) & \numprint{0} (0-5.7k) \\
	
  \hline
	
 All & 1.1m & \numprint{6897} & 402m & 20k (1k-1.2m) & 2 (1-15k) & 40 (1-507k) & \numprint{0} (0-8.7k) & \numprint{0} (0-19.2k) \\

\end{tabular}
\end{table*}

\section{Study Results}
\label{sec:study-results}

This section presents study results. It is structured according to the research questions and summarizes the findings in boxes.

\subsection{General Descriptive Statistics}

Table~\ref{tab:code-quality-overview} presents descriptive statistics for the quality metrics of the analysed repositories. It shows median values where \textit{higher} percentage values indicate \textit{lower} quality. Clone coverage ranges between 7\% and 11\%, indicating that on average about 1/10 of the code is part of at least one clone. There are no big differences between programming languages. Comment incompleteness ranges between 60\% and 75\%, meaning that on average almost 3/4 of public code entities are not commented. Java code exhibits notably higher quality in that respect than code written in C, C++ or C\#. Please note that this metric was not computed for JavaScript repositories. 

The median of the metric ``too long files" is 37\%, indicating that roughly one-third of the code is located in files which are longer than 750 source lines of code. This metric varies a lot between programming languages, namely from 10\% (Java) to 70\% (JavaScript). The median of the metric ``too long methods" is 18\%, indicating that on average one-fifth of the code is placed in methods whose length exceed 75 source lines of code. Metric values for programming languages vary a lot from 3\% (JavaScript) to 31\% (C). Nesting depth ranges between 1\% and 5\%, indicating that on average very few code is located in deeply nested methods. We summarize the descriptive findings in the following statements:

\begin{leftbar}
\begin{itemize}
\item
Only a small fraction of code bases is part of a clone or located in deeply nested methods. For these metrics, there is no strong difference between C, C++, C\#, Java and JavaScript code.

\item
3/4 of public code entities are not documented by a comment. Java code is documented more completely than code written in C, C++, or C\#.

\item
1/3 of the code is located in long files. This value heavily varies between programming languages, ranging from 9\% for Java code to 70\% for JavaScript code. 

\item
1/5 of the code is located in long methods. This value varies between programming languages, ranging from 3\% for JavaScript code to 30\% for C code.
\end{itemize}
\end{leftbar}


\begin{table}
  \centering
\caption[Overview of Metric Results]{Overview of Metric Results, Median Values, Higher \% values indicate lower quality, ``--": No data}
\label{tab:code-quality-overview}
\begin{tabular}{l r r r r r}
                  & Clone       & Comment          & File   & Method & Nesting \\
 Language & Coverage & Incompleteness & Size & Length  & Depth\\
  \hline
	C & 7\% & 72\% & 48\% & 31\% & 5\% \\
	C++ & 8\% & 74\% & 36\% & 21\% & 4\% \\
	C\# & 11\% & 75\% & 16\% & 13\% & 3\% \\
	Java & 9\% & 60\% & 10\% & 10\% & 2\% \\
	JavaScript & 9\% & -- & 70\% & 3\% & 1\% \\
	\hline
  All & 8\% & 72\% & 37\% & 18\% & 3\% \\
	\hline
\end{tabular}
\end{table}

\subsection{How does the programming language affect maintainability (RQ1)?}

\noindent\textbf{H1: C code has more code duplication than code written in other languages.}\\
Table~\ref{tab:code-quality-overview} shows the median clone coverage values by language. These figures already suggest that there is only little difference between C-projects and the ones in other languages. The overall null hypothesis that there is no difference between the code duplication
in code of different languages has to be rejected however (p-value $<0.0001$). Hence, we can look in more detail into the comparison
of C code against other code. The pairwise comparisons show that the null hypotheses that there is no difference in clone coverage between C code and the
respective other languages cannot be rejected (p-value of 1 for all comparisons, d between 0.04 and 0.22).

\begin{leftbar}
C code has no higher clone coverage than code written in C++, C\#, Java, or JavaScript.
\end{leftbar}

\noindent\textbf{H2: Code written in C\# and Java has no differences regarding maintainability.}\\
Here, we only analysed the data from C\# and Java projects. The MANOVA analysis showed that we have to reject the null hypothesis that
there is no difference in the quality of C\# and Java code (p-value$<$0.0001). In the ANOVA analyses of single metrics, we found only in 
nesting depth that there was no significant difference (p-value=0.38, d=0.05). In the Wilcoxon tests, significant differences are in clone coverage (p-value =0.0006, d=0.19), large files (p-value$<$0.0001, d=0.34), large methods (p-value$<$0.0001, d=0.22) and comment completeness (p-value$<$0.0001). Hence, we have to reject the hypothesis that there is
no difference in quality between C\# and Java code apart from their nesting depth. This also fits to the median values in Tab.~\ref{tab:code-quality-overview} which show
better values for Java then for C\# for all metrics. Yet, the effect sizes are small.

\begin{leftbar}
Java code shows better maintainability than C\# code.
\end{leftbar}
%

\noindent\textbf{H3: JavaScript code is more deeply nested than code written in other languages.}\\
The Kruskal-Wallis test showed that we need to reject the null hypotheses that there is no difference between the deep nesting of code in 
different programming languages (p-value$<$0.0001). Hence, we can go beyond the omnibus hypothesis
and look more closely on JavaScript.  The pairwise comparisons of the Wilcoxon test showed that we cannot reject the hypothesis that
there is no difference in nesting depth between JavaScript and any of the other four language (p-value = 1 for all languages, d between 0.01 and 0.34). In fact, the median
nesting depth is lower than in all other languages.


\begin{leftbar}
JavaScript code does not differ in deep nesting from code written in C, C++, C\#, or Java.
\end{leftbar}

\noindent\textbf{H4: C code has more very long methods than code written in other languages.}\\
The Kruskal-Wallis test showed that the null hypothesis that there is no difference in the number of long methods between code in different
languages has to be rejected (p-value$<$ 0.0001). Hence, we can look further in the specific position of C. All pairwise Wilcoxon tests showed
that the null hypotheses has to be rejected (p-value$<$0.0001 for all languages, d=0.44 for C++, d=0.73 for C\#, d=0.94 for Java, d=0.99 for JavaScript). We have to accept the alternative hypothesis that C code has
more very long methods than code written in other languages.

\begin{leftbar}
C code has longer methods than code written in C++, C\#, Java, or JavaScript.
\end{leftbar}
 
\noindent\textbf{H5: The quality of interface documentation is similar for all languages.}\\
Since ConQAT does not support the detection of JavaScript documentation, we limited the scope of this analysis to the four remaining languages.
The Kruskal-Wallis test showed that the null hypothesis that there is no difference in the quality of interface documentation across languages
has to be rejected (p-value$<$0.0001). A further analysis using pairwise Wilcoxon tests showed statistically significant differences only between 
Java and C ($\delta =0.25$), C++ ($\delta =0.25$) and C\# ($\delta =0.22$, p-values all $<$0.0001). There is no statistically significant difference between C, C++, C\# and JavaScript as well as between Java and JavaScript (all p-values= 1, $\delta$s between 0.01 and 0.31). Hence, we have to reject the hypothesis that
the quality of interface documentation is similar for all languages. Instead, we have support for a new hypothesis that the quality of interface documentation
is better in Java code then in most other code.

\begin{leftbar}
The interface documentation of Java code is better than for code written in C, C++, and C\# and there is no difference between the three latter languages.
\end{leftbar}
%

\noindent\textbf{H6: JavaScript code has the lowest maintainability among all languages.}\\
The null hypothesis that there is no difference in maintainability among code bases written in different languages has to be rejected.
The MANOVA analysis gave a p-value smaller than 0.0001. A further look into the single ANOVA analyses showed that this holds for
all single used quality metrics. Hence, we analyzed the pairwise comparisons with JavaScript using Wilcoxon tests. We analyzed the nesting depth
already in H3 and found no difference. We also already investigated comment completeness in H5 and found no difference. The further analysis 
showed no significant difference
in clone coverage (C: p-value=0.03, d=0.29, C++: p-value=0.35, d=0.26, C\#: p-value=1, d=0.05 and Java:
p-value=1, d=0.22). For large files, JavaScript is significantly worse than all other languages with medium to large effect sizes (d=0.60 for C, d=1.04 for
C++, d=1.54 for C\# and d=1.99 for Java, p-values $<$0.0001 for all languages). Finally, for long methods,
there is no statistically significant difference (p-value=1 for all languages, d between 0.12 and 0.91). The large effect sizes here are because the
JavaScript tends to have the lowest number of too long methods. Based on this, with only one quality metric in which JavaScript is worst,
we decided to reject the hypothesis that JavaScript
code has the lowest maintainability in all analyzed languages. Instead, we adopt the new hypothesis that there is no overall difference in quality
of code written in the analysed languages.

\begin{leftbar}
Maintainability of JavaScript code is \emph{not} lower compared to code written in C, C++, C\#, or Java.
\end{leftbar}

\subsection{Which non-language aspects influence or are influenced by maintainability (RQ2)?}
\label{sec:non-language-aspects}

\noindent\textbf{H7: Code developed by a team has better quality than code developed by an individual}\\
The null hypotheses that there is no difference in the quality of code developed by different numbers of contributors could not be
rejected. The MANOVA analysis gave a p-value of 0.26. Hence, we cannot accept H7. Code developed in teams is not associated with
higher maintainability.

\begin{leftbar}
Code developed in teams does not have better quality then code developed by an individual.
\end{leftbar}

\noindent\textbf{H8: Large code bases have a lower quality than small code bases.}\\
The MANOVA analysis showed that the null hypothesis that there is no difference in quality of code bases of different sizes has to be
rejected (p-value$<$0.0001). Looking into the individual ANOVA analyses, only the metric ``too long files'' showed a difference.
To investigate this in more detail, we calculated Pearson's product-moment correlation with correlation coefficient 0.1004 and a p-value$<$0.0001.
Hence, large code bases tend to have more very large files. Yet, there is only a difference for this one metric with a small effect size.
Therefore, we propose that we cannot accept H8. Low maintainability is not more associated with larger code bases.

\begin{leftbar}
Large code bases do not have worse maintainability then small code bases.
\end{leftbar}

\noindent\textbf{H9: Repositories with high maintainability have more forks than repositories with low maintainability.}\\
The null hypotheses that there is no difference in the number of forks in repositories with different qualities cannot be rejected. The ANOVA
analysis gave p-values of 0.12 (clone coverage), 0.91 (nesting depth), 0.25 (large files), 0.97 (large methods) and 0.90 (comment 
completeness). Hence, we cannot accept H9. High maintainability is not associated with more forks.

\begin{leftbar}
Repositories with high maintainability do not have more forks.
\end{leftbar}

\noindent\textbf{H10: Repositories with high maintainability are more popular than repositories with low maintainability.}\\
The null hypotheses that there is no difference in popularity in repositories with different qualities cannot be rejected. The ANOVA
analysis gave p-values of 0.021 (clone coverage), 0.56 (nesting depth), 0.19 (large files), 0.84 (large methods) and 0.91 (comment 
completeness). Hence, we cannot accept H10. High maintainability 
of projects is not associated with more popularity.

\begin{leftbar}
Repositories with high maintainability are not more popular.
\end{leftbar}

\section{Discussion}
\label{sec:discussion}

This section discusses study results, summarizes implications, and presents threats to validity. 

\subsection{Results Discussion}
\label{sec:results-discussion}

When analyzing the resulting metric values (cf. Table~\ref{tab:code-quality-overview}), the question arises whether they are ``good" or ``bad". As all metric values are percentages, they indicate the likelihood of encountering a code smell, e.g.\ a clone, when selecting an arbitrary line from the code base. The higher this likelihood, the more often developers will have to deal with such maintainability issues. The median values for clone coverage and nesting depth are low, indicating that open-source developers rarely have to struggle with clones or deeply-nested methods. The rather low clone coverage values for code written in C and C++ match results by Koschke and Bazrafshan~\cite{koschke2016software}. 

In contrast, half of the C code and 2/3 of JavaScript code is located in long files, which implies that C and JavaScript developers have to cope with long files frequently. Finally, 72\% of all public code entities in code bases written in C, C++, C\# or Java are not documented. Hence, developers frequently have to read and understand those in detail, e.g.\ for review or reuse purposes, which costs time and productivity and could be avoided.

Our results show that Java code has on average the highest and C code the lowest maintainability. Apart from the programming language, there might be other reasons to explain this finding: differing education on and community support for maintainability, differing refactoring support by IDEs, or the use of static analysis tools. We are not able to investigate these aspects and their impact on maintainability and leave this for future work.

Overall, only a single hypothesis could be supported by our large sample of open-source repositories: C code has longer methods than code written in
the other languages. As C is the only non-object-oriented language in our analysis, we can see this as indication that structuring code according to object-oriented principles helps in writing smaller methods. All other hypotheses related to programming languages could not be supported.

To further study the impact of programming languages on maintainability, we identified top repositories by intersecting the sets of the 25\% best repositories for all five metrics. Likewise, we identified flop repositories. Interestingly, repositories from each language are among the top and flop repositories, indicating that is is possible to write high-quality -- but also low-quality -- code in every language. Hence, we conclude that  language has a small impact on maintainability.

This result is in agreement with Ray et al.~\cite{Ray2014large,Ray2017} and Bissyand\'e et al.~\cite{bissyande2013popularity}. Summarizing, three ways of operationalizing quality -- bug fix commits, the number of issues, and static analysis metrics -- come to the same conclusion: The programming language has likely only a modest influence on maintainability.

Not one hypothesis on maintainability and development activity, repository popularity, code base size, or team size could be supported. These results contrast with related work which found that software quality is influenced by the size of the code base~\cite{elemam2001confounding} or the number of developers~\cite{bird2011dont}. But they are in accordance with Weyuker et al.~\cite{weyuker2008do} and Norick et al.~\cite{norick2010effects}, who found that the number of developers has no major impact on code quality. Additionally, they agree with Ahmed et al.~\cite{ahmed2014exploration} who found that code quality is not affected by code base size. In contrast, Ray et al. ~\cite{Ray2014large} found a strong correlation between the number of commits and code quality. This difference can be explained that their operationalization of quality (the number of bug fix commits) grows proportionally with the number of commits while ours does not. Furthermore, they are similar to results by Corral et al.~\cite{corral2015better} who found that the code quality of Android apps has only a marginal impact on market success.

\subsection{Implications for Researchers and Practitioners}

\subsubsection*{Implications for Researchers}


That only 1 of 10 hypotheses was accepted shows the importance of empirical research to test assumptions about maintainability. 
When comparing study results with related work, we found that software quality is operationalized differently in different studies, e.g. as number of bug fix commits, number of issues, number of post-release defects, number of pre-release defects, or static analysis results. Hence, we find it interesting to compare these quality measurements -- e.g.\ study whether they are correlated or if results are stable when changing the operationalization -- and suggest future work in this direction. Furthermore, researchers should study contradictions between results of different studies regarding relationships of non-language aspects with maintainability.

\vspace{-1.4em}
\subsubsection*{Implications for Practitioners}

Most importantly, practitioners should take the results of this study to challenge their own conceptions about maintainability. One might argue that the hypotheses
we formulate do not represent such prejudices well. Yet, we show that all such conceptions need to be empirically tested if they are used for project decisions.

Furthermore, practitioners can use the metric values presented in this paper as a reference when interpreting static analysis results from their own code base. If a metric value for an own code base is worse than the average, this demonstrates that it is easily possible to write better code and it might encourage practitioners to improve their code base.

Moreover, practitioners should investigate whether their own code base suffers from problems and consider taking countermeasures. C developers should especially look at comment incompleteness, file size, and method length. C++ developers should in particular regard comment incompleteness and file size. C\# and Java developers should especially look at comment incompleteness. JavaScript developers should, in particular, be aware of file sizes. 

\subsection{Threats to Validity}


As we considered a random sample of a huge size from GitHub as the de-facto standard for open-source hosting today, we are rather confident that the results generalize to
open-source systems in the investigated languages.

As we randomly sampled repositories from GitHub, the majority of repositories is rather small~\cite{Kalliamvakou2014promises} and this might bias the results. To address this threat, we used a minimum threshold of \numprint{10000} lines of code for the size of a repository.

Before performing static analysis, we excluded irrelevant code automatically (cf. Section~\ref{sec:data-collection-procedures}). Due to the heterogeneity of GitHub repositories, we might not have excluded all irrelevant code which might bias the results. To minimize this effect, we improved the approach developed by H\"onick\cite{Hoenick2015does} and manually verified the absence of irrelevant code in a random sample.

We did not discriminate between formatting styles when calculating the sizes of files and methods. Formatting styles denote whether opening and closing brackets are placed on separate lines. This fact might bias results to the disadvantage of C\# code  where brackets are usually put on separate lines. To address this threat, we chose high threshold values for size metrics.

When analyzing comment incompleteness, we targeted public code entities. While this concept is rather clear for code written in C\#, Java, and JavaScript (because the languages provide visibility features), it is not so clear for code written in C and C++. For C, we considered all function definitions in header files. For C++, we considered the definitions of classes, methods and attributes in header files. This operationalization might bias the results when developers document code entities outside header files.

\section{Conclusion}
\label{sec:conclusion}

This paper presented a large-scale empirical study on prejudices about maintainability that we evaluated on open-source code bases. We used a random sample of \numprint{6897} GitHub repositories containing 402 million lines of code written in C, C++, C\#, Java and JavaScript. We automatically excluded irrelevant code and used static analysis to determine the maintainability of each code base. Based on this information, we investigated the impact of programming languages and other factors on maintainability. We provide all information necessary to replicate the study on GitHub.
In agreement with related studies, we found that the programming language has only a modest impact on maintainability. In addition, we found that there is no significant relationship between maintainability and development activity, repository popularity, code base size and team size. This indicates that these factors and the programming language are no decisive factors regarding maintainability. 

Future work should investigate first, what reasons and motivations are behind the results and elicit best practices for maintainability. To this end, we plan to compare top and flop GitHub repositories and interview developers. Second, it should evaluate the effects of code base size on the results. Third, it should consider more languages and more unproven assumptions. Fourth, it should study differences between open-source and closed-source code. Fifth, it should compare different operationalizations of maintainability.

\balance

%
%
 \bibliographystyle{splncs04}
 \bibliography{paper-bibliography}

\end{document}